# An analysis of voltage source inverter switches fault classification using short time Fourier transform

**Mustafa Manap[1], Srete Nikolovski[2], Aleksandr Skamyin[3], Rony Karim[4], Tole Sutikno[5], Mohd Hatta Jopri[6]**

[1,6]Faculty of Electrical & Electronic Engineering Technology, Universiti Teknikal Malaysia Melaka, Malacca, Malaysia
[2]Power Engineering Department, Faculty of Electrical Engineering, Computer Science and Information Technology, University of Osijek, 31000 Osijek, Croatia
[3]Department of Electric Power and Electromechanics, Saint Petersburg Mining University, Saint Petersburg, Russia
[4]Rohm Semiconductor (ROHM) GmbH, Germany
[5]Deparment of Electrical Engineering, Universitas Ahmad Dahlan, Yogyakarta, Indonesia
[5]Embedded System and Power Electronics Research Group (ESPERG), Yogyakarta, Indonesia

| Article Info | ABSTRACT |
|---|---|
| *Article history:*<br><br>Received Jul 10, 2021<br>Revised Oct 9, 2021<br>Accepted Oct 21, 2021<br><br>*Keywords:*<br><br>Open circuit fault<br>Rule-based classifier<br>Short circuit fault<br>Short time Fourier transform<br>Voltage source inverter | The dependability of power electronics systems, such as three-phase inverters, is critical in a variety of applications. Different types of failures that occur in an inverter circuit might affect system operation and raise the entire cost of the manufacturing process. As a result, detecting and identifying inverter problems for such devices is critical in industry. This study presents the short-time Fourier transform (STFT) for fault classification and identification in three-phase type, voltage source inverter (VSI) switches. Time-frequency representation (TFR) represents the signal analysis of STFT, which includes total harmonic distortion, instantaneous RMS current, RMS fundamental current, total non harmonic distortion, total waveform distortion and average current. The features of the faults are used with a rule-based classifier based on the signal parameters to categorise and detect the switch faults. The suggested method's performance is evaluated using 60 signals containing short and open circuit faults with varying characteristics for each switch in VSI. The classification results demonstrate the proposed technique is good to be implemented for VSI switches faults classification, with an accuracy classification rate of 98.3%.<br><br> |

*Corresponding Author:*

Mustafa Manap
Faculty of Electrical & Electronic Engineering Technology
Universiti Teknikal Malaysia Melaka (UTeM)
Jalan Hang Tuah Jaya, 76100 Durian Tunggal, Melaka, Malaysia
Email: mustafa@utem.edu.my

## 1. INTRODUCTION

Voltage source inverter (VSI) is commonly used to convert a DC voltage into an AC voltage with varying magnitude and frequency [1]–[6]. As a single failure in the converter components will result in a fault in the overall system, the system reliability of these converters is critical [7]–[15]. As a result, in several crucial procedures, the VSI must be able to operate continuously even under malfunctioning situations [16]–[19]. Semiconductor switches and aluminium capacitors are the two most critical components of a voltage source inverter. Soldering joint and semiconductor failures account for more than 34% of all malfunctions and failures [20]–[24].

The voltage supply inverter, like any other power electronics energy conversion device, is directly exposed to excessive stress elements such as physical stress and thermal, electrical [25]–[29]. Power semiconductor switches, particularly and insulated gate bipolar transistors (IGBTs), and meta-oxide semiconductor field-effect transistors (MOSFETs) are the most sensitive components in the VSI, resulting in greater failure rates [30]–[35]. Power switch faults are divided into two types: short circuit faults (SCFs), and open circuit faults (OCFs) [36]–[41]. Moreover, OCF arises as a result of driver failure, which retracts the bonding wires during thermic cycling by a SCF caused by rupture or age [42]–[45]. OCF poses risk to the key components of a energy transfer, and healthy converter to load is often achieved even in a deteriorated state [46]–[48]. However, if such failures persist for an extended period of time, more damage to the converter may occur, resulting in an utter stop in the worst-case scenario [49]–[53]. As a result, identifying and detecting such failures is suggested to avoid severe damage in power converters [54]–[58].

Numerous studies have proposed various methods for three-phase inverter open circuit (OC) failure diagnostics, including model-based algorithms and signal processing-based algorithms especially time-frequency domain analysis [31], [59]–[65]. Several researchers have presented model-based methods for diagnosing faulty switches through all the study of the system model, which demonstrate excellent accuracy and strong applicability but require a precise mathematical model or additional hardware [12], [66]–[72]. Previous model-based techniques, on the other hand, are susceptible to system characteristics and have limited detection accuracy [73]–[75]. Prior knowledge regarding converter specifications (parasitic resistances, inductances, capacitances) is critical for the improvement of model-based algorithms, but difficult to gain from the existing system [16], [44], [76]. Shahbazi *et al.* [49] proposed a model-based approach for detection of fault using a field-programmable gate array (FPGA) based on time and voltage criteria. Nevertheless, when compared to an application specific integrated circuit (ASIC) platform with the same design specifications, the FPGA does not provide fast execution, reduced power consumption, or lower mass manufacturing costs [1], [77]–[84]. Additionally, Jiang *et. al* and Lei *et. al* stated in [85], [86] that the typical convolution neural network (CNN) analysis for OC had a flaw with the model-based algorithm since the required the size of the sample data used for model training was considerable and prone to overfitting.

There is no requirement to find prior information of converter parameters in this proposed method, thus a development of signal processing fault diagnostic method is presented. The proposed method, known as the short-time fourier transform (STFT), is based on time-frequency domain analysis [87], [88]. The suggested method will discover OC faults in converter topologies by classifying specified signal parameters. Furthermore, combining STFT analysis with a ruled-based classifier may overcome the limitations of earlier techniques by providing fast detection, simple implementation, and high accuracy.

## 2. RESEARCH METHOD

Figure 1 shows the flow chart of the proposed method. The fault signals are modelled using MATLAB software and represented in both frequency, and time domain namely the STFT. From the TFR, the significant signal parameters such as instantaneous of total waveform distortion, RMS current, total nonharmonic distortion, average current, total harmonic distortion and RMS fundamental current are estimated. The switches fault are then classified and identified using a rule-based classifier based on the signal characteristics.

### 2.1. Modeling of VSI switches fault

Despite the fact that overcurrent or short circuit protection against power switches has become increasingly popular on industrial drives, open circuit failures are still disregarded in the industrial context [89]. Overheating from thermic cycling, for example, can cause connections to break, which can be caused by bonding wire lifting, device driver problems, or connection breakage itself [90]. Furthermore, the OC has the ability to develop secondary faults and cause several device problems in the system if used in such an unusual manner over an extended period of time [91]. The topology of open and short circuit faults of VSI are shown in Figure 2(a) and 2(b), recpectively. Figure 3(a) and 3(b) depict a VSI switches fault model using MATLAB. In addition, the designed circuit has a DC voltage input of 50V, a sampling time of 100s, and a fundamental frequency of 60 Hz.

### 2.2. Short time Fourier transform

STFT is a significant tool for indicating a signal in both the frequency and time [84], [92]–[94]. The signal spectral characteristic in time domain can be recognised using the TFR. As a result, STFTs are an appropriate method for analysing switch fault signals with non-stationary and multi-frequency components.

$$S_x(t,f) = \int_{-\infty}^{\infty} x(t)w(\tau-t)e^{-j2\pi ft}d\tau \tag{1}$$

where ω is the observation window.

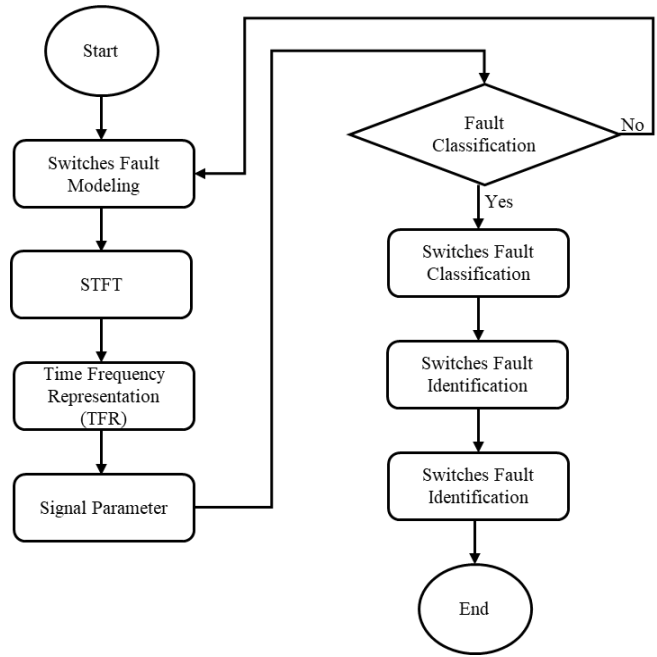

Figure 1. Flowchart of the proposed method

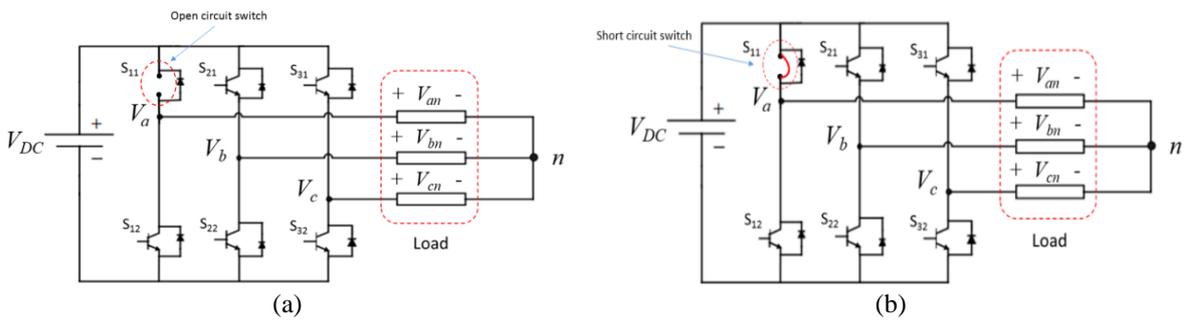

Figure 2. The fault topology of VSI, (a) Open circuit switch fault at leg 'A', (b) short circuit switch fault at leg 'A'

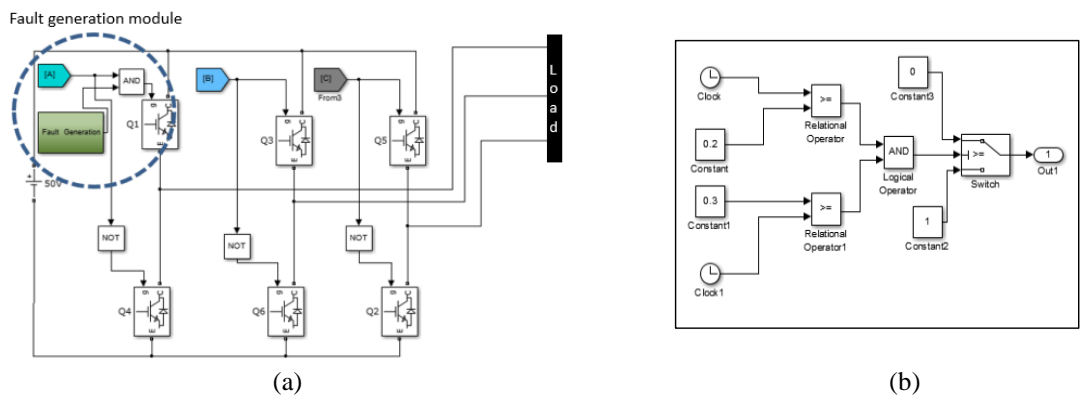

Figure 3. VSI switches fault model using MATLAB, (a) Switches fault model at leg A, (b) Fault generation module

## 2.3. Signal characteristic

The STFT is used to calculate signal characteristics in order to offer timely data about the signal. It is distinguished by the following parameters: total harmonic distortion mean (THD$_{mean}$), fault duration (T$_{d,fault}$), total non-harmonic distortion mean (TnHD$_{mean}$), average RMS current mean (I$_{rms,mean}$), average current mean (I$_{ave,mean}$), and total waveform distortion mean (TWD$_{mean}$). Using signal parameter information, it is then feasible to identify the faulty switch location. Furthermore, the sample frequency for the recommended approach is 10 kHz.

## 2.4. Ruled-based classifier

A deterministic classification, which is rule-based classifier, is widely utilised in a variety of applications due to its simplicity and ease of implementation [95]. The efficiency of the classification is strongly depending on competent threshold, and rules preferences. To classify and identify failures in the switches, a rule-based classifier based on signal characteristics is used. This study will use 60 different signals with varied characteristics of different switch fault signal in order to establish the optimum potential threshold setup parameters for the suggested technique. In addition, the total waveform distortion threshold (TWD$_{thres,fault}$) is used to calculate the threshold values of the proposed method.

$$T_{d,fault} = \int_0^T \begin{cases} 1, for\ TWD(t) \geq TWD_{thres,fault} \\ 0, x \geq elsewhere \end{cases} \quad (2)$$

## 3. RESULTS AND DISCUSSION

This section discusses the findings of the switch faults analysis using STFT. The open switch fault take place at S$_{31}$ of phase c upper, and the short switch fault appears at S$_{12}$ of phase. Figure 4(a) and 4(b) depict the time and time-frequency domains of open circuit fault signals. Figure 4(a) illustrates the signal in TFR, where the red colour represents the largest magnitude and the blue colour represents the lowest magnitude, as seen in the contour plots. Meanwhile, Figure 4(b) depicts the magnitude of the signal as a transient reduce from 1.6A to 0.05A for a fault period of 60 ms during an open circuit fault.

Following that, a short-circuit fault of the VSI switch occurs at S$_{12}$ and. As demonstrated in 60 Hz frequency appears along the time axis in Figure 5(a). The DC component (0 Hz) does exist, however, during the fault period, which ranges from 195ms to 255 ms. Figure 5(b) shows the signal magnitude abruptly drops from 1.6 A to 0.05 A for 60 ms between 195 ms and 255 ms.

Figure 6 depicts a graphical representation of the average current, fundamental current, and RMS current of open circuit switch faults. Starting at 195 ms and lasting 60 ms, the RMS current is reduced from 1.17 A to 0.9 A. The basic current exhibits a similar behaviour, with the signal's current suddenly dropping for 60 ms. The current decline from the nominal value ranges from 1.17 to 0.7 A. Similarly, once the fault signals are recognised, the average current signal is reduced to -0.7 A. As illustrated in Figure 7(a), the RMS current for a short circuit increases abruptly to 1.35 A from the nominal value which is 1.17 A, whereas for a period of 60 ms the RMS fundamental current declines to 0.75 A. Comparable to average current, the current drops from 0 A to -1.1 A and exists at the negative cycle.

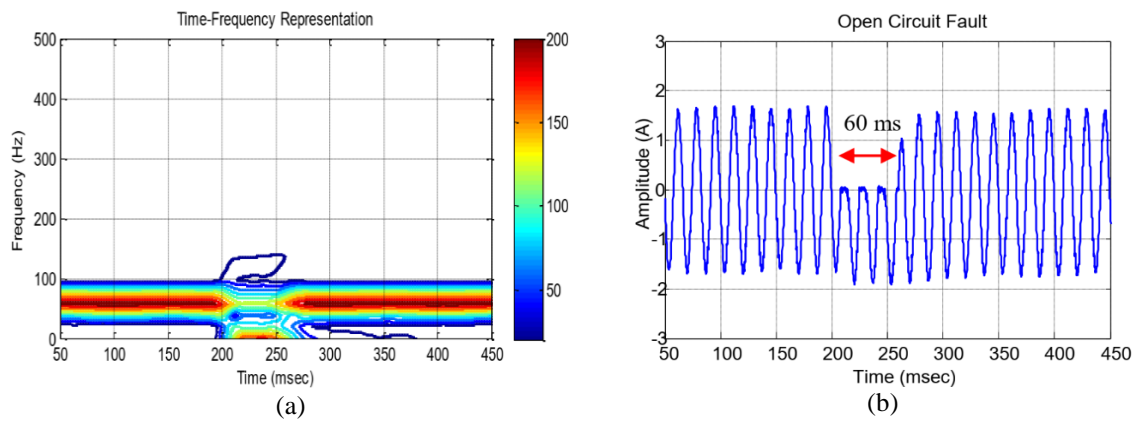

Figure 4. Switch fault at S$_{31}$, (a) time-frequency representation (TFR), (b) open circuit fault signals

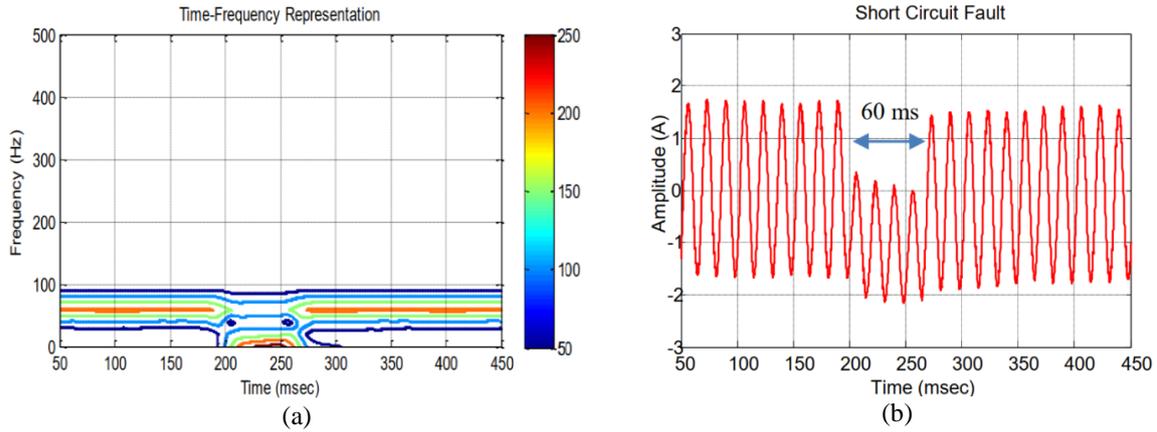

Figure 5. Switch fault at $S_{12}$ (a) short circuit fault signals, (b) time-frequency representation (TFR)

Figures 8(a), and 8(b) present the total waveform distortion and total nonharmonic distortion magnitude for a period 60 ms that increases from 2% to 55% and 47%, respectively. As shown in Figure 8(c), the magnitude of total harmonic distortion increases by 30% for open circuit switches fault. According to the investigation of open circuit switch faults, total nonharmonic distortion has a greater percentage value than total harmonic distortion. Total waveform distortion is the combination of total nonharmonic distortion, and total harmonic distortion.

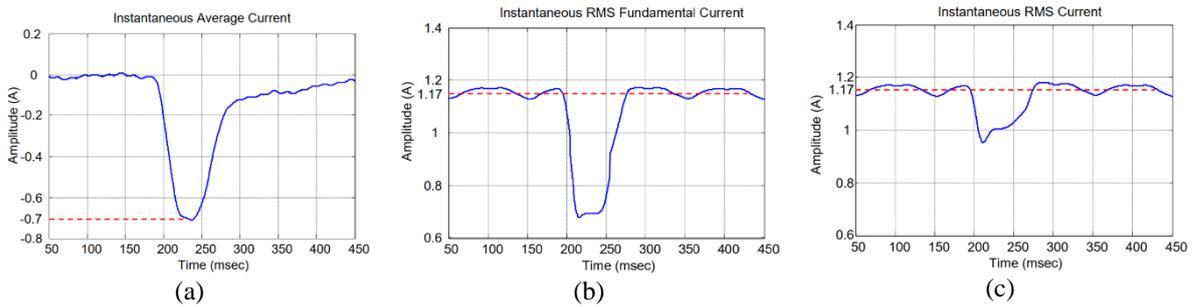

Figure 6. open circuit switches faults signal parameter; (a) instantaneous average current, (b) instantaneous fundamental current and (c) instantaneous rms current.

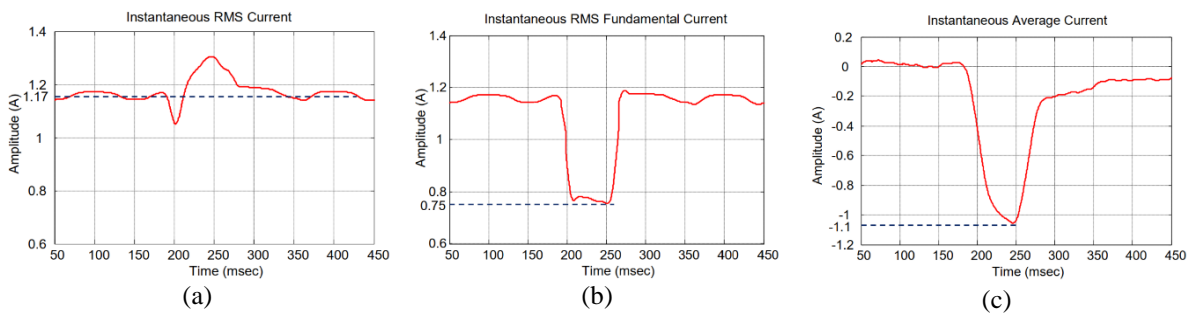

Figure 7. short circuit switches faults signal parameter; (a) instantaneous rms current, (b) instantaneous fundamental current and (c) instantaneous average current

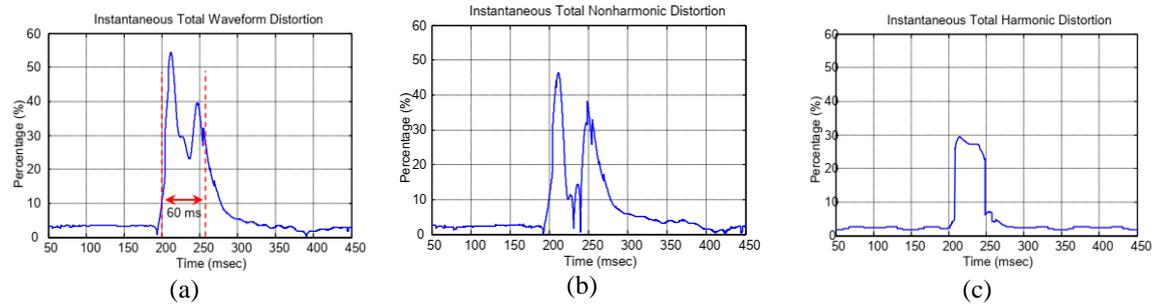

Figure 8. signal parameters of open circuit switches faults; (a) instantaneous of total waveform distortion, (b) total nonharmonic distortion, and (c) total harmonic distortion

Figure 9 (a) depicts the amount of total waveform distortion or short circuit switches fault, which is identical to total nonharmonic distortion since total harmonic distortion and total nonharmonic distortion are added together. Meanwhile, as seen in Figure 9, the amount of total harmonic distortion remains low (3%). As a result of this, Figure 9 (c) depicts the total nonharmonic distortion magnitude, which increases from 4% to 55% at 180 ms for a period of 60 ms.

The fault signal was derived through the analysis of 60 signals with varied properties for each type of switch (open and short circuit for $S_{11}$, $S_{12}$, $S_{21}$, $S_{22}$, $S_{31}$, and $S_{32}$). The best threshold value for a rule-based classifier is found to be 0.05 or 5%. The pseudo code as shown in Figure 10 describes a rule-based classifier for classifying and identifying switch faults based on signal characteristics. The fault signals of switches are evaluated and classified using STFT and a rule-based method. In addition, 60 signals with various fault signal characteristics are generated to determine the system's performance. According to Table 1, the proposed technique provides 98.3% accuracy of classification.

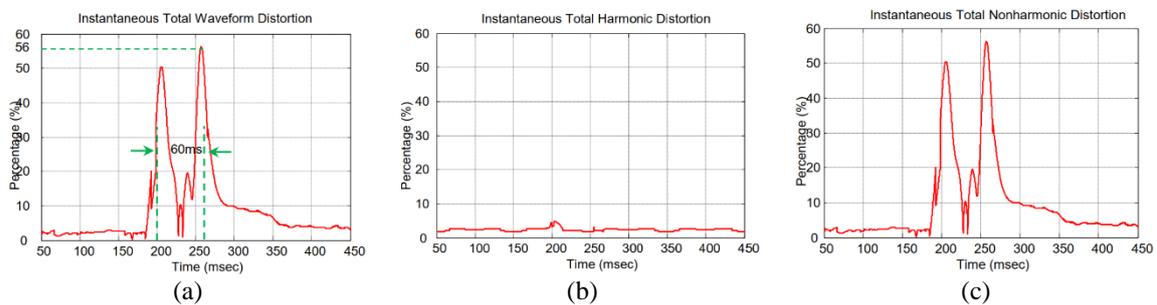

Figure 9. Short Circuit Switches Faults Signal Parameter; (a) Instantaneous Total Waveform Distortion, (b) Total Harmonic Distortion and (c) Total Nonharmonic Distortion

```
Fucntion[z]=rule_basic_classifier (T_{d,fault},
I_{ave,mean}, I_{rms,mean}, THD_{ave,mean}, TnHD_{ave,mean}, TWD_{ave,mean})
If(011011)
z = Open: S1;
else if (100111)
z = Open: S1;
else if (100111)
z = Open: S2;
else if (011111)
z = Open: S2;
else if (101101)
z = Open: S3;
else if (0100101)
z = Open: S3;
else if (010101)
z = Open: S4;
else if (101010)
z = Open: S4;
else if (110110)
z = Open: S5;
else if (001011)
z = Open: S6;
else if (110011)
z = Short: S6;
else
z = unknown
end
```

Figure 10. Pseudo code of a rule-based classifier for classifying and identifying switch faults

Table 1. Performance of switches faults signals classification of the proposed method

| Faults | Switch | % Correct Classification |
|---|---|---|
| Open-Circuit | S1 | 98.3 |
| | S2 | 98.3 |
| | S3 | 98.3 |
| | S4 | 98.3 |
| | S5 | 98.3 |
| | S6 | 98.3 |
| Short-Circuit | S1 | 98.3 |
| | S2 | 98.3 |
| | S3 | 98.3 |
| | S4 | 98.3 |
| | S5 | 98.3 |
| | S6 | 98.3 |

**4. CONCLUSION**

The STFT approach, in conjunction with a rule-based classifier, was used to create a classification and identification system for switch fault signals. The system's performance is then validated by categorising 60 actual signals with varying characteristics for each sort of switch fault signal. The suggested approach performs admirably, with 98.3 percent of faults correctly classified. As a result, it shows that the system is well-suited for usage as a switch fault monitoring system. Other time-frequency domain methods, such as the Gabor transform and S-transform, should be investigated in future study to improve classification accuracy.

**ACKNOWLEDGMENTS**

This research is supported by the Advanced Digital Signal Processing Laboratory (ADSP Lab). Special thanks also to the Faculty of Electrical & Electronic Engineering Technology of Universiti Teknikal Malaysia Melaka (UTeM), Center for Robotics and Industrial Automation (CeRIA) of UTeM and Ministry of Higher Education Malaysia (MOHE). Their support is gratefully acknowledged. The author's also thanks to University of Osijek, Saint Petersburg Mining University, Rohm Semiconductor, Universitas Ahmad Dahlan, and Embedded System and Power Electronics Research Group for supporting this collaborative research in the present work.

**BIOGRAPHIES OF AUTHORS**

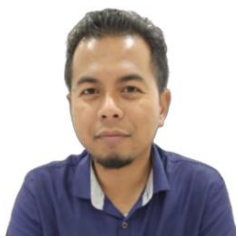

**Mustafa Manap** was born in Kuala Lumpur, Malaysia on 1978. He received his B.Sc from Universiti Technologi Malaysia in 2000 and Msc. in Electrical Engineering from Universiti Teknikal Malaysia Melaka (UTeM) 2016. Since 2006, he has been an academia staff at Universiti Teknikal Malaysia Melaka (UTeM). He is registered with Malaysia Board of Technologist (MBOT), and a member of International Association of Engineers (IAENG). His research interests are power electronics and drive, instrumentation, and DSP application.

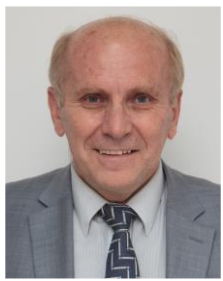

**Srete Nikolovski** is a tenured full professor and senior advisor in tenure at the Faculty of Electrical Engineering in Osijek. In 1989 he received his MSc degree from the Faculty of Electrical Engineering in Belgrade. In 1990 he started to work at the Faculty of Electrical Engineering in Osijek. November 30, 1993, he earned his doctoral dissertation at the FER, University of Zagreb. He became a full professor in 2000. During his work in the Department of Power Engineering, he has published 5 coursebooks. He was a member of the executive board of HRO Cigre. He is also an IEEE Senior member of the Power Engineering Society, Reliability Engineering Society. He published over 250 papers in International Journals and Conferences. He was a supervisor of more than 300 graduate student works, 6 MSc thesis, and 9 Ph.D. student dissertations. He is a senior IEEE member with papers in IEEE Transaction of SG, IEEE System Journal, IEEE Journal of Emerging and Selected Topics in Power Electronics.

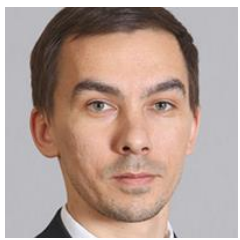
**Aleksandr Skamyin** was born in Volkhov, Russia on 1986. He received his B.Eng and Ph.D. degree in electrical engineering from Saint-Petersburg Mining University (Russian Federation) in 2007 and 2011, respectively. From 2011 till 2015 he was the assistant in National Mineral Resources University (Mining University), Saint Petersburg, Russian Federation. Since 2015, he is an associated professor in Saint-Petersburg Mining University (Russian Federation). His research interests include power quality, reactive power compensation in the presence of harmonics, regulation of power consumption mode.

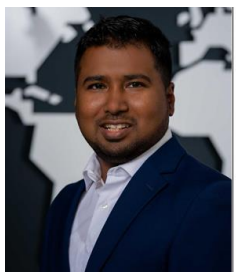
**Rony Karim** was born in Bogra, Bangladesh on 1985. He received his B.Engg from Rajshahi University of Engineering & Technology (RUET) in 2006, Msc. in Electrical Power Engineering from Rheinisch-Westfälische Technische Hochschule Aachen (RWTH), Germany in 2011 and in addition, he has completed MBA from HHL Leipzig Graduate School of Management, Germany in 2019, and he is a member of International Association of Engineers (IAENG), respectively. He is an experienced power electronics engineer and Since 2011, he has been serving well-known Multinational semiconductor companies and held various engineering, sales & marketing positions with in semiconductor industry in Europe. His focus areas are Industrial Drives, Factory automation and power trains for e-mobility. Currently he is serving ROHM semiconductor GmbH, Germany as an Application Marketing Manager for European Region.

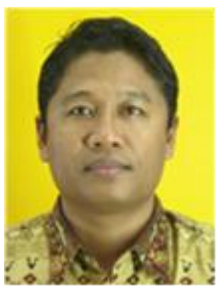
**Tole Sutikno** is a Lecturer in Electrical Engineering Department at the Universitad Ahmad Dahlan (UAD), Yogyakarta, Indonesia. He received his B.Eng., M.Eng. and Ph.D. degrees in Electrical Engineering from Universitas Diponegoro, Universitas Gadjah Mada and Universiti Teknologi Malaysia, in 1999, 2004 and 2016, respectively. He has been an Associate Professor in UAD, Yogyakarta, Indonesia since 2008. He is currently an Editor-in-Chief of the TELKOMNIKA since 2005, and the Head of the Embedded Systems and Power Electronics Research Group since 2016. His research interests include the field of industrial applications, industrial electronics, industrial informatics, power electronics, motor drives, renewable energy, FPGA applications, embedded system, image processing, artificial intelligence, intelligent control, digital design and digital library.

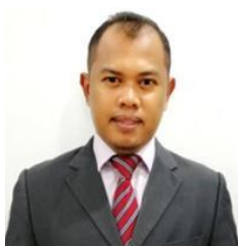
**Mohd Hatta Jopri** was born in Johor, Malaysia on 1978. He received his B.Eng from Universiti Teknologi Malaysia (UTM), Msc. in Electrical Power Engineering from Rheinisch-Westfälische Technische Hochschule Aachen (RWTH), Germany, and PhD from Universiti Teknikal Malaysia Melaka (UTeM), respectively. Since 2005, he is an academia and research staff at UTeM. He is registered with Malaysia Board of Technologist (MBOT), and a member of International Association of Engineers (IAENG). His research interests include power electronics and drive, power quality analysis, signal processing, machine learning and data science.